\documentclass{article}
\usepackage{amssymb}

\usepackage{graphicx}
\usepackage{amsmath}

\input{tcilatex} 

\begin{document}

\title{\textbf{Light-front gauge propagator reexamined}}
\author{Alfredo T.Suzuki and J.H.O.Sales \\
%EndAName
Instituto de F\'{\i}sica Te\'{o}rica, 01405-900 S\~{a}o Paulo, Brazil.}
\maketitle

\begin{abstract}
Gauge fields are special in the sense that they are invariant under gauge
transformations and \emph{``ipso facto''} they lead to problems when we try
quantizing them straightforwardly. To circumvent this problem we need to
specify a gauge condition to fix the gauge so that the fields that are
connected by gauge invariance are not overcounted in the process of
quantization. The usual way we do this in the light-front is through the
introduction of a Lagrange multiplier, $(n\cdot A)^2$, where $n_\mu$ is the
external light-like vector, i.e., $n^2=0$, and $A_\mu$ is the vector
potential. This leads to the usual light-front propagator with all the
ensuing characteristics such as the prominent $(k\cdot n)^{-1}$ pole which
has been the subject of much research. However, it has been for long
recognized that this procedure is incomplete in that there remains a
residual gauge freedom still to be fixed by some ``ad hoc'' prescription,
and this is normally worked out to remedy some unwieldy aspect that emerges
along the way. In this work we propose a new Lagrange multiplier for the
light-front gauge that leads to the correctly defined propagator with no
residual gauge freedom left. This is accomplished via $(n\cdot A)(\partial
\cdot A)$ term in the Lagrangian density. This leads to a well-defined and
exact though Lorentz non invariant propagator.
\end{abstract}

\section{Introduction}

The history of the light-front gauge goes as far back as 1949 with the
pioneering work of P.A.M.Dirac \cite{dirac}, where the front-form of
relativistic dynamics was introduced as a well-defined possibility for
describing relativistic fields. Since its d\'ebut into quantum field theory
it has known days of both glory and oblivion for varied reasons. On the one
hand it seemed a solid grounded and more convenient approach to studying
quantum fields, e.g., the only setting where a proof of the finiteness of
the $N=4$ supersymmetric Yang-Mills theory could be carried out successfully
was in the light-cone gauge (a facet of its glory) \cite{susylc}. But on the
other hand, manifest Lorentz covariance is lost and non-local terms sneak
into the renormalization program (the other side of the coin that charges us
with a price to pay).

One of the reasons why the light-front form has lured many into this field
of research is due to the fact that its propagator structure seemed simple
enough to deserve their special attention. However, its manifest apparent
simplicity hide many complexities not envisaged at first glance nor
understood without much hard work. For example, one of the, say, ``ugly''
aspects of the ensuing propagator is the emergence of the mistakenly
so-called ``unphysical'' pole which in any physical processes of interest
leads to Feynman integrals bearing these singularities. We say mistakenly
because as it became understood later, it is in fact very much physical in
that without a proper treatment of such a pole, one violates basic physical
principles such as causality \cite{pimentelsuzuki}.

On the other hand, for the brighter side of it, the light-front gauge seemed
advantageous in quantum field theory because it allowed the possibility of
decoupling the ghost fields in the non-Abelian theories, since it is an
axial type gauge, as shown by J. Frenkel \cite{josif}, a property that can
simplify Ward-Takahashi identities \cite{WTI} and problems involving
operator mixing or diagram summation \cite{opmix}.

Looking through the light-front literature we soon realize that there is a
simple and standard gauge vector potential field propagator in which appears
two terms \cite{Kogut}, namely, 
\begin{equation}  \label{prop}
G^{\mu\nu}_{ab}(k) = \frac {-i\delta ^{ab}}{k^2}\left \{g^{\mu\nu}-\frac{
k^\mu n^\nu+k^\nu n^\mu)}{k\cdot n}\right \}\,,
\end{equation}
where $a, b$ labels non-Abelian gauge group indices.

We see that the propagator (\ref{prop}) has one strictly covariant factor
proportional to the space-time metric $g^{\mu \nu }$ and also the
characteristic light-front factor proportional to $(k^{\mu }n^{\nu }+k^{\nu
}n^{\mu })(k\cdot n)^{-1}$. For the majority of computations, be they in
quantum field theory or in nuclear physics (Bethe-Salpeter, etc.) make use
of this propagator. Some people have recognized the presence of a third term
proportional to $(k^{2}n^{\mu }n^{\nu })(k\cdot n)^{-2}$ \cite{hari}, i.e., 
\begin{equation}
G_{ab}^{\mu \nu }(k)=\frac{-i\delta ^{ab}}{k^{2}}\left\{ g^{\mu \nu }-\frac{
k^{\mu }n^{\nu }+k^{\nu }n^{\mu }}{k\cdot n}+\frac{k^{2}n^{\mu }n^{\nu }}{
(k\cdot n)^{2}}\right\} ,  \label{correct}
\end{equation}
but this third term has always been consistently dropped in the actual
calculations on the grounds that it has been claimed long ago that such 
\emph{``contact terms''} have no physical significance because they do not
propagate any information. After all, from its inception, the paradigm has
always been that gauge terms such as $k^{\mu }n^{\nu }+k^{\nu }n^{\mu }$ and 
$k^{2}n^{\mu }n^{\nu }$ must not contribute to any physical process because
of current conservation. If that be the case, then we must squarely face the
vexing question: Why one would drop only the \emph{``contact terms''} in the
calculations on the grounds that they do not have physical significance
because propagates no information? However, more recently, it has been shown 
\cite{jorgehenrique} that this is not the case. These \emph{``contact terms''
} do have physical significance being carriers of relevant information.

Our contribution in this paper is to show that the condition $n\cdot A=0$ ($
n^2=0$) is \emph{necessary} but \emph{not sufficient} to define the
light-front gauge. It leads to the standard form of the light-front
propagator (\ref{prop}) which lacks the relevant contact term of (\ref
{correct}). The \emph{necessary} and \emph{sufficient} condition to uniquely
define the light-front gauge is given by $n\cdot A=\partial \cdot A=0$ so
that the corresponding Lagrange multiplier to be added to the Lagrangian
density is proportional to $(n\cdot A)(\partial \cdot A)$ instead of the
usual $(n\cdot A)^2$. Note that the condition $\partial \cdot A = 0$ in the
light-cone variables defines exactly (for $n\cdot A=A^+=0$) the constraint 
\begin{equation*}
A^-=\frac{\partial^{\perp}A^{\perp}}{\partial ^{+}} \Rightarrow \frac{
k^{\perp}A^{\perp}}{k^+}.
\end{equation*}
This constraint, together with $A^+=0$, once substituted into the Lagrangian
density yields the so-called two-component formalism in the light-front,
where one is left with only physical degrees of freedom, and Ward-Takahashi
identities and multiplicative renormalizability of pure Yang-Mills field
theory is verified \cite{gluonvertex}. Thus, if we start off by correctly
defining the gauge condition in the light-front form, the problems related
to residual gauge freedom, zero modes and ..... are completely finessed. 
%checar verbo finesse. 

\section{Light-Front Dynamics: Definition}

According to Dirac \cite{dirac} it is \emph{``...the three-dimensional
surface in space-time formed by a plane wave front advancing with the
velocity of light. Such a surface will be called front for brevity ''}. An
example of a light-front is given by the equation $x^{+}=x^{0}+x^{3}$.

A dynamical system is characterized by ten fundamental quantities: energy,
momentum, angular momentum, and boost. In the conventional Hamiltonian form
of dynamics one works with dynamical variables referring to physical
conditions at some instant of time, the simplest instant being given by $
x^{0}=0 $. Dirac found that other forms of relativistic dynamics variables
refer to physical conditions on a front $x^{+}=0$. The resulting dynamics is
called light-front dynamics, which Dirac called front-form for brevity.

A perusal into the specific literature will soon help us to discover that
many different names are used to describe this form of dynamics and the
corresponding gauge, such as light-front field theory, field theory in the
infinite momentum frame, null plane field theory and light-cone field
theory. We prefer the word light-front since the quantization surface is a
light-front (tangential to the light cone).

The variables $x^{+}=x^{0}+x^{3}$ and $x^{-}=x^{0}-x^{3}$ are called
light-front \emph{``time''} and \emph{longitudinal space} variables
respectively. Transverse variables are $x^{\perp }=(x^{1},x^{2})$. We call
the reader's attention to the fact that there are many different conventions
used in the literature. Here, we follow the conventions, notations and some
useful relations employed in \cite{Kogut}.

By analogy with the light-front space-time variables, we define the
longitudinal momentum $k^{+}=k^{0}+k^{3}$ and light-front \emph{``energy''} $
k^{-}=k^{0}-k^{3}$.

For a free massive particle, the on-shell condition $k^{2}=m^{2}$ leads to $
k^{+}\geq 0$ and the dispersion relation 
\begin{equation}  \label{disprel}
k^{-}=\frac{(k^{\perp })^{2}+m^{2}}{k^{+}}.
\end{equation}

This dispersion relation (\ref{disprel}) is quite remarkable for the
following reasons: \emph{(1)} Even though we have a relativistic dispersion
relation, there is no square root factor. \emph{(2)} The dependence of the
energy $k^{-}$ on the transverse momentum $k^{\perp }$ is just like in the
nonrelativistic relation. \emph{(3)} For $k^{+}$ positive (negative), $k^{-}$
is positive (negative). This fact has several interesting consequences. 
\emph{(4)} The dependence of energy on $k^{\perp }$ and $k^{+}$ is
mutiplicative and large energy can result from large $k^{\perp }$ and/or
small $k^{+}$. This simple observation has drastic consequences for
renormalization aspects \cite[4]{wilson90}

\section{Massless vector field propagator}

The Lagrangian density for the vector gauge field (for simplicity we
consider an Abelian case) is given by 
\begin{equation}
\mathcal{L}=-\frac{1}{4}F_{\mu \nu }F^{\mu \nu }-\frac{1}{2}\left( \partial
_{\mu }A^{\mu }\right) ^{2},  \label{1}
\end{equation}
where the characteristic Lagrange multiplier proportional to $(\partial
\cdot A)^{2}$ is the so-called gauge-breaking term, which defines a physical
configuration space, the space of orbits, from projecting out the gauge
fields onto this space.

The equations of motion in the light-front variables are 
\begin{equation}
\partial ^{+}\left[ \frac{1}{2}\partial ^{+}A^{-}+\frac{1}{2}\partial
^{-}A^{+}-\partial ^{\perp }A^{\perp }\right] -\left( \partial ^{+}\partial
^{-}-\partial ^{\perp 2}\right) A^{+}=0  \label{2}
\end{equation}
\begin{equation}
\partial ^{j}\left[ \frac{1}{2}\partial ^{+}A^{-}+\frac{1}{2}\partial
^{-}A^{+}-\partial ^{\perp }A^{\perp }\right] -\left( \partial ^{+}\partial
^{-}-\partial ^{\perp 2}\right) A^{j}=0  \label{3}
\end{equation}
\begin{equation}
\partial ^{-}\left[ \frac{1}{2}\partial ^{+}A^{-}+\frac{1}{2}\partial
^{-}A^{+}-\partial ^{\perp }A^{\perp }\right] -\left( \partial ^{+}\partial
^{-}-\partial ^{\perp 2}\right) A^{-}=0  \label{4}
\end{equation}

The usual procedure in the light-front \emph{milieu} has been to make a
gauge choice by taking \cite{Kogut},\cite{Neville} 
\begin{equation}
A^{+}=0\,.  \label{5}
\end{equation}

This gauge choice is known as infinite-momentum gauge, null-plane gauge,
light-cone gauge and light-front gauge. From (\ref{2}), we have 
\begin{equation}
\partial ^{+}A^{-}=2\partial ^{\perp }A^{\perp }+F(x^{+},x^{\perp })
\label{6}
\end{equation}
Thus $A^{-}$ is not a dynamical variable. Choosing $F$ to be zero, the
dynamical variables $A^{i}$ obey the massless Klein-Gordon equation.

Since the dynamical variables $A^{j}$ obey massless Klein-Gordon linear
equation, the general solution will be given by the superposition of plane
waves: 
\begin{equation}
A^{j}(x)=\int \frac{dk^{+}d^{2}k^{\perp }}{2k^{+}(2\pi )^{3}}\sum_{\alpha=1,2}\delta _{j\alpha }\left[ a_{\alpha }(k)e^{-ikx}+a_{\alpha }^{\dagger
}(k)e^{ikx}\right] .  \label{6.1}
\end{equation}
The operator $a_{\alpha }(k)$ and $a_{\alpha }^{\dagger }(k)$ are
annihilation and creation operators for photons. They satisfy the
commutation relations 
\begin{eqnarray}
\left[ a_{\alpha }(k),a_{\beta }^{\dagger }(k^{\prime })\right] &=&2(2\pi
)^{3}k^{+}\delta _{\alpha \beta }\delta ^{3}(k-k^{\prime })  \notag \\
\left[ a_{\alpha }(k),a_{\beta }(k^{\prime })\right] &=&0\text{ \ , \ }\left[
a_{\alpha }^{\dagger }(k),a_{\beta }^{\dagger }(k^{\prime })\right] =0.
\label{6.2}
\end{eqnarray}
The equal $x^{+}$ commutation relation for the transverse components
of the gauge field is
\begin{equation}
\left[ A^{j}(x),A^{l}(y)\right] _{x^{+}=y^{+}}=\frac{-i}{4}\delta
_{jl}\:\epsilon (x^{-}-y^{-})\delta ^{2}(x^{\perp }-y^{\perp }),  \label{6.3}
\end{equation}
where the indices $j,\,l$ label transverse components of the field.

Taking into consideration the commutators among the field operators as
derived above, we may write the momentum space expansions of the free field
operator. Introducing the polarization vectors \cite{prem} 
\begin{equation}
\epsilon _{1}^{\mu }(k)=\frac{1}{k^{+}}(0,2k^{1},k^{+},0)\text{, \ }\epsilon
_{2}^{\mu }(k)=\frac{1}{k^{+}}(0,2k^{2},0,k^{+}),  \label{6.4}
\end{equation}
we can write 
\begin{equation}
A^{\mu }(x)=\int \frac{dk^{+}d^{2}k^{\perp }}{2k^{+}(2\pi )^{3}}\sum_{\alpha
}\epsilon _{\alpha }^{\mu }(k)\left[ a_{\alpha }(k)e^{-ikx}+a_{\alpha
}^{\dagger }(k)e^{ikx}\right] .  \label{6.5}
\end{equation}
We obtain the Lorentz condition (note that this is \emph{a posteriori}
condition) 
\begin{equation}
\partial \cdot A=0.  \label{6.7}
\end{equation}
Introducing the four-vector $n=(1,0,0,-1)$ we have the relation 
\begin{equation}
\sum_{\alpha=1,2}\epsilon _{\alpha }^{\mu }(k)\epsilon _{\alpha }^{\nu
}(k)=-g^{\mu \nu }+\frac{n^{\mu }k^{\nu }+k^{\mu }n^{\nu }}{k^{+}}-n^{\mu
}n^{\nu }\frac{k^{2}}{(k^{+})^{2}}.  \label{6.8}
\end{equation}

Let $G^{\mu \nu }=iS^{\mu \nu }$ denote the massless vector field propagator 
\cite{yan} in the light-front theory. We have 
\begin{eqnarray}
S^{\mu \nu }(x-y) &=&-i\left\langle 0\right| T^{+}A^{\mu }(x)A^{\nu
}(y)\left| 0\right\rangle  \notag \\
G^{\mu \nu }(x-y) &=&\theta (x^{+}-y^{+})\left\langle 0\right| A^{\mu
}(x)A^{\nu }(y)\left| 0\right\rangle +  \notag \\
&&+\theta (y^{+}-x^{+})\left\langle 0\right| A^{\nu }(y)A^{\mu }(x)\left|
0\right\rangle .  \label{6.9}
\end{eqnarray}
Using the expansion (\ref{6.5}) we have 
\begin{equation*}
G^{\mu \nu }(x-y)=\int \frac{dk^{+}d^{2}k^{\perp }}{2k^{+}(2\pi )^{3}}\frac{
e^{-ik(x-y)}}{k^{2}+i\varepsilon }\left[ -g^{\mu \nu }+\frac{n^{\mu }k^{\nu
}+k^{\mu }n^{\nu }}{k^{+}}-n^{\mu }n^{\nu }\frac{k^{2}}{(k^{+})^{2}}.\right]
\end{equation*}

\section{Propagator with gauge fixing $(n\cdot A)(\partial \cdot A)=0$}

In this (and in the subsequent appendices) instead of going through
the canonical procedure of determining the propagator as done in the
previous section, we shall adopt a more head-on, classical procedure
by looking for the inverse operator corresponding to the differential
operator sandwiched between the vector potentials in the Lagrangian
density.

The relevant gauge fixing term that enters in the Lagrangian density
we define as
\begin{equation}
(n\cdot A)(\partial \cdot A)=0,
\end{equation}
yielding for the Abelian gauge field Lagrangian density: 
\begin{equation}
\mathcal{L} =-\frac{1}{4}F_{\mu \nu }F^{\mu \nu }-\frac{1}{2\alpha }\left(
2 n_\mu A^{\mu}\partial _{\nu }A^{\nu }\right)=\mathcal{L}_{\text{E}}+\mathcal{L}_{GF}
\label{8}
\end{equation}
where the gauge fixing term is conveniently written so as to
symmetrize the indices $\mu$ and $\nu$. By partial integration and
considering that terms which bear a total derivative don't contribute
and that surface terms vanish since $ \lim\limits_{x\rightarrow \infty
}A^{\mu }(x)=0$, we have
\begin{equation}
\mathcal{L}_{\text{E}}=\frac{1}{2}A^{\mu }\left( \square g_{\mu \nu
}-\partial _{\mu }\partial _{\nu }\right) A^{\nu }  \label{9}
\end{equation}
and 
\begin{equation}
\mathcal{L}_{GF}=-\frac{1}{\alpha }(n\cdot A)(\partial\cdot A)=-\frac{1}{
2\alpha }A^{\mu } \left( n_{\mu }\partial _{\nu }+n_{\nu }\partial _{\mu
}\right) A^{\nu }  \label{25}
\end{equation}
so that 
\begin{equation}
\mathcal{L} =\frac{1}{2}A^{\mu }\left( \square g_{\mu \nu }-\partial
_{\mu }\partial _{\nu }-\frac{1}{\alpha }(n_\mu\partial _{\nu }+n_\nu
\partial _{\mu })\right) A^{\nu } \label{11}
\end{equation}

To find the gauge field propagator we need to find the inverse of the
operator between parenthesis in (\ref{11}). That differential operator in
momentum space is given by:
\begin{equation}
O_{\mu \nu }(k)=-k^{2}g_{\mu \nu }+k_{\mu }k_{\nu }+\frac{1}{\alpha }\left(
n_{\mu }k_{\nu }+n_{\nu }k_{\mu }\right)  \label{op}
\end{equation}
so that the propagator of the field, which we call $G^{\mu \nu }(k)$, must
satisfy the following equation: 
\begin{equation}
O_{\mu \nu }G^{\nu \lambda }\left( k\right) =\delta _{\mu }^{\lambda }
\label{12}
\end{equation}

$G^{\nu \lambda }(k)$ can now be constructed from the most general
tensor structure that can be defined, i.e., all the possible linear
combinations of the tensor elements that composes it (the most general form includes the light-like vector $m_{\mu }$ dual to the
$n_{\mu }$ \cite{progress} -- but for our present
purpose it is in fact indifferent): 
\begin{eqnarray}
G^{\mu \nu }(k) &=&g^{\mu \nu }A+k^{\mu }k^{\nu }B+k^{\mu }n^{\nu }C+n^{\mu
}k^{\nu }D+k^{\mu }m^{\nu }E+  \notag \\
&&+m^{\mu }k^{\nu }F+n^{\mu }n^{\nu }G+m^{\mu }m^{\nu }H+n^{\mu }m^{\nu
}I+m^{\mu }n^{\nu }J  \label{a2}
\end{eqnarray}
Since (\ref{op}) does not contain any $m_{\mu }$ factors it is
straightforward to conclude that $E=F=H=I=J=0$. Then, we have 
\begin{equation}
A=-(k^{2})^{-1}  \label{a3}
\end{equation}
\begin{equation}
A+(k\cdot n)C+\lambda (k\cdot n)B+\lambda n^{2}C=0  \label{a5}
\end{equation}
\begin{equation}
-k^{2}C+\lambda A+\lambda k^{2}B+\lambda (k\cdot n)C=0  \label{a6}
\end{equation}
\begin{equation}
(k\cdot n)G+\lambda A+\lambda (k\cdot n)D+\lambda n^{2}G=0  \label{a7}
\end{equation}
\begin{equation}
-k^{2}G+\lambda k^{2}D+\lambda (k\cdot n)G=0  \label{a9}
\end{equation}
where $\lambda \equiv \alpha ^{-1}$.

From (\ref{a9}) we have 
\begin{equation*}  \label{a10}
G=\frac{\lambda k^{2}}{(k^{2}-\lambda k\cdot n)}D,
\end{equation*}
which inserted into (\ref{a7}) yields 
\begin{eqnarray}
D&=&\frac{(k^{2}-\lambda k\cdot n)}{\left[ \lambda (k\cdot
n)^{2}-2k^{2}k\cdot n-\lambda k^{2}n^{2}\right] }A,  \notag \\
D&=&\frac{(\alpha k^2-k\cdot n)}{\left[(k\cdot n)^{2}-2\alpha k^{2}k\cdot
n-k^{2}n^{2}\right] }A,
\end{eqnarray}
so that substituting $D$ back in (\ref{a10}) gives 
\begin{eqnarray}
G&=&\frac{\lambda k^{2}}{\left[ \lambda (k\cdot n)^{2}-2k^{2}k\cdot
n-\lambda k^{2}n^{2}\right] }A.  \notag \\
G&=&\frac{k^2}{(k\cdot n)^2-2\alpha k^2 k\cdot n-k^2n^2}A.
\end{eqnarray}

From (\ref{a5}) and (\ref{a6}) 
\begin{equation*}
\left\{ 
\begin{array}{c}
A+\lambda (k\cdot n)B+(k\cdot n+\lambda n^{2})C=0 \\ 
\lambda A+\lambda k^{2}B+(-k+\lambda k\cdot n)C=0
\end{array}
\right. ,
\end{equation*}
obtaining for the system 
\begin{equation*}
C=\frac{(\alpha k^{2}-k\cdot n)}{\left[(k\cdot n)^{2}-2\alpha k^{2}k\cdot
n-k^{2}n^{2}\right] }A=D
\end{equation*}
and 
\begin{equation*}
B=\frac{(\alpha^2 k^{2}+n^{2})}{\left[(k\cdot n)^{2}-2\alpha k^{2}k\cdot
n-k^{2}n^{2}\right] }A.
\end{equation*}

In the light-front $n^{2}=0$ and taking the limit $\alpha \rightarrow 0$, we
have 
\begin{eqnarray}
B&=&0\,,  \notag \\
C&=&-\frac{A}{k\cdot n}  \notag \\
D&=&-\frac{A}{k\cdot n}  \notag \\
G&=&\frac{k^{2}}{(k\cdot n)^{2}}A.
\end{eqnarray}

Then, it is a matter of straightforward algebraic manipulation to get
the relevant propagator in the light-front gauge, namely,
\begin{equation}
G^{\mu \nu }(k)=-\frac{1}{k^{2}}\left\{ g^{\mu \nu }-\frac{k^{\mu }n^{\nu
}+n^{\mu }k^{\nu }}{k\cdot n}+\frac{n^{\mu }n^{\nu }}{(k\cdot n)^{2}}
k^{2}\right\} \,,  \label{27}
\end{equation}
which has the outstanding third term commonly referred to as \emph{contact
term}.

\section{Conclusions}

We have constructed a Lagrange multiplier in the light-front that completely
fixes the gauge choice so that no unphysical degrees of freedom are left. In
other words, no residual gauge remains to be dealt with. Moreover this
allows us to get the correct propagator including the important contact
term. As have been proved, this term is of capital importance in the
renormalization of (Bethe-Salpeter?) ...

The configuration space wherein the gauge potential $A_\mu$ is defined have
by the gauge symmetry many equivalent points for which we can draw an
immaginary line linking them. These constitute the gauge potential orbits.
Gauge fixing therefore means to select a particular orbit. The light-front
condition $n\cdot A=0$ defines a hypersurface in the configuration space
which cuts the orbits of the gauge potentials. This surface is not enough to
completely fix the gauge. We also need the hypersurface $\partial \cdot A=0$
. The intersect between the two hypersurfaces defines a clear cut line and a
preferred direction in the configuration space. The two together then
completely fixes the gauge with no residual gauge freedom left.

\vspace{.5cm}

{\sc acknowledements:} {\tt A.T.Suzuki is partially supported by CNPq under process 303848/2002-2 and J.H.O.Sales is supported by FAPESP under process 00/09018-0}

\vspace{.5cm}

\appendix

\section{Propagator with gauge fixing $\partial \cdot A=0$} 

The gauge fixing term known as Lorentz condition 
\begin{equation}
\partial \cdot A=0,  \label{7}
\end{equation}
yields for the Abelian gauge field Lagrangian density: 
\begin{equation}
\mathcal{L} =-\frac{1}{4}F_{\mu \nu }F^{\mu \nu }-\frac{1}{2\alpha }\left(
\partial _{\mu }A^{\mu }\right) ^{2}=\mathcal{L}_{\text{E}}+\mathcal{L}_{GF}
\label{8}
\end{equation}

By partial integration and considering that terms which bear a total
derivative don't contribute and that surface terms vanish since $
\lim\limits_{x\rightarrow \infty }A^{\mu }(x)=0$, we have 
\begin{equation}
\mathcal{L}_{\text{E}}=\frac{1}{2}A^{\mu }\left( \square g_{\mu \nu
}-\partial _{\mu }\partial _{\nu }\right) A^{\nu }  \label{9}
\end{equation}
and 
\begin{equation}
\mathcal{L}_{GF}=-\frac{1}{2\alpha }\partial _{\mu }A^{\mu }\partial _{\nu
}A^{\nu }=\frac{1}{2\alpha }A^{\mu }\partial _{\mu }\partial _{\nu }A^{\nu }
\label{10}
\end{equation}
so that 
\begin{equation}
\mathcal{L} =\frac{1}{2}A^{\mu }\left( \square g_{\mu \nu }-\partial _{\mu
}\partial _{\nu }+\frac{1}{\alpha }\partial _{\mu }\partial _{\nu }\right)
A^{\nu }  \label{11c}
\end{equation}

To find the gauge field propagator we need to find the inverse of the
operator between parenthesis in (\ref{11c}). That differential operator in
momentum space is given by: 
\begin{equation}
O_{\mu \nu }=-k^{2}g_{\mu \nu }+k_{\mu }k_{\nu }-\frac{1}{\alpha }k_{\mu
}k_{\nu }\,,  \label{11a}
\end{equation}
so that the propagator of the field, which we call $G^{\mu \nu }(k)$, must
satisfy the following equation: 
\begin{equation}
O_{\mu \nu }G^{\nu \lambda }\left( k\right) =\delta _{\mu }^{\lambda }
\label{12}
\end{equation}

$G^{\nu \lambda }(k)$ can now be constructed from the most general tensor
structure that can be defined, i.e., all the possible linear combinations of
the tensor elements that composes it: 
\begin{equation}
G^{\nu \lambda }(k)=\left[ Ag^{\nu \lambda }+Bk^{\nu }k^{\lambda }+Cn^{\nu
}n^{\lambda }+Dk^{\nu }n^{\lambda }+Ek^{\lambda }n^{\nu }\right]  \label{13}
\end{equation}
where $A$, $B$, $C$, $D$ and $E$ are coefficients that must be determined in
such a way as to satisfy (\ref{12}). Of course, it is immediately clear that
since (\ref{11}) does not contain any external light-like vector $n_{\mu }$,
the coefficients $C=D=E=0$ straightaway. So, 
\begin{equation}
G^{\mu \nu }(k)=-\frac{1}{k^{2}}\left\{ g^{\mu \nu }-(1-\alpha )\frac{k^{\mu
}k^{\nu }}{k^{2}}\right\}  \label{14}
\end{equation}

Of course, this is the usual covariant Lorentz gauge, which for $\alpha=1$
is known as Feynman gauge and for $\alpha=0$ as Landau gauge.

\section{Propagator with gauge fixing $n\cdot A=0$}

The axial type gauge fixing is accomplished through the condition 
\begin{equation}
n_{\mu }A^{\mu }=0  \label{15}
\end{equation}
so that we can write the Lagrangian density as 
\begin{equation}
\mathcal{L} =-\frac{1}{4}F_{\mu \nu }F^{\mu \nu }-\frac{1}{2\alpha }\left(
n_{\mu }A^{\mu }\right) ^{2}=\mathcal{L}_{\text{E}}+\mathcal{L}_{GF}
\label{16}
\end{equation}

In a similar way as before, we have: 
\begin{equation}
\mathcal{L}_{\text{E}}=\frac{1}{2}A^{\mu }\left( \square g_{\mu \nu
}-\partial _{\mu }\partial _{\nu }\right) A^{\nu }  \label{17}
\end{equation}
and 
\begin{equation}
\mathcal{L}_{GF}=-\frac{1}{2\alpha }n_{\mu }A^{\mu }n_{\nu }A^{\nu }=-\frac{1
}{2\alpha }A^{\mu }n_{\mu }n_{\nu }A^{\nu }.  \label{18}
\end{equation}
Therefore 
\begin{equation}
\mathcal{L} =\frac{1}{2}A^{\mu }\left( \square g_{\mu \nu }-\partial _{\mu
}\partial _{\nu }-\frac{1}{\alpha }n_{\mu }n_{\nu }\right) A^{\nu }
\label{19}
\end{equation}

In momentum space the relevant differential operator that needs to be
inverted is given by 
\begin{equation}
O_{\mu \nu}=-k^{2}g_{\mu \nu }+k_{\mu }k_{\nu }-\frac{1}{\alpha }n_{\mu
}n_{\nu }\,,  \label{20}
\end{equation}
so that, the general tensorial structure given in (\ref{13}) that must
satisfy (\ref{12}) yields 
\begin{equation}
G^{\mu \nu}(k)=-\frac{1}{k^{2}}\left\{ g^{\mu \nu }-\frac{k^{\mu }k^{\nu }}{\left( k\cdot n\right) ^{2}}\left( n^{2}-\alpha k^{2}\right) -\frac{k^{\mu}n^{\nu }+n^{\mu }k^{\nu }}{k\cdot n}\right\} \,\,.  \label{23b}
\end{equation}
\begin{equation}
G^{\mu \nu }(k)=-\frac{1}{k^{2}}\left\{ g^{\mu \nu }-\frac{k^{\mu }k^{\nu }}{
\left( k\cdot n\right) ^{2}}\left( n^{2}-\alpha k^{2}\right) -\frac{k^{\mu
}n^{\nu }+n^{\mu }k^{\nu }}{k\cdot n}\right\} \,\,.  \label{23}
\end{equation}

Taking the limit $\alpha \rightarrow 0$ and using the light-like vector $
n_{\mu }$ for which $n^{2}=0$ we have finally 
\begin{equation}
G^{\mu \nu }(k)=-\frac{1}{k^{2}}\left[ g^{\mu \nu }-\frac{\left( k^{\mu
}n^{\nu }+n^{\mu }k^{\nu }\right) }{\left( k\cdot n\right) }\right] \,,
\label{24}
\end{equation}
which is the standard two-term light-front propagator so commonly found in
the literature.


\begin{thebibliography}{99}
\bibitem{dirac}  P.A.M.Dirac, Rev.Mod.Phys.\textbf{21}, 392 (1949).

-*\bibitem{susylc}  J.H.Schwarz, Phys.Rep.\textbf{89}, 223 (1982);
S.Mandelstam, Nucl.Phys.\textbf{B213} 183 (1983); L.Brink \emph{et al},
Phys.Lett.\textbf{B123}, 323 (1983).

\bibitem{pimentelsuzuki}  B.M.Pimentel and A.T.Suzuki, Phys.Rev.\textbf{D42}
, 2115 (1990).

\bibitem{josif}  J.Frenkel, \emph{Theories in Algebraic Non-covariant Gauges}
, World Scientific Publishing Co. Ltd. (1991).

\bibitem{WTI}  D.M.Capper and D.R.T.Jones, Phys.Rev. \textbf{D31}, 3295
(1985).

\bibitem{opmix}  C.B.Thorn, Phys.Rev.\textbf{D20} 1934 (1979); D.J.Pritchard
and W.J.Stirling, Nucl.Phys.\textbf{B165}, 237 (1980).

\bibitem{Kogut}  J.B.Kogut and D.E.Soper, Phys.Rev.\textbf{D1} (1970) 2901.

\bibitem{hari}  T.-M.Yan, Phys.Rev.\textbf{D7}, 1780 (1973); A.Harindranath, 
\emph{Light-Front Quantization and Non-Perturbative QCD}, editors J.P.Vary
and F.W\"{o}lz, International Institute of Theoretical and Applied Physics
(1997).

\bibitem{jorgehenrique} J.H.O.Sales, Tobias Frederico, and
B.M.Pimentel, Hadrons Review 2002.

\bibitem{gluonvertex}  A.T.Suzuki, Z.Phys.\textbf{C38}, 595 (1988).

\bibitem{wilson90}  S.D.Glasek and K.G.Wilson, Phys.Rev.\textbf{D49} (1994)
4214; J.H.O. Sales, T.Frederico, B.V. Carlson and P.U. Sauer, Phys. Rev. 
\textbf{C63}:064003(2001).

\bibitem{Neville}  R.A.Neville and F.Rohrlich, Phys.Rev.\textbf{D3} (1971)
1692.

\bibitem{prem}  Prem P.Srivastava and Stanley J.Brodsky, Phy.Rev.\textbf{D66}
:045019 (2002)

\bibitem{yan}  S.-J. Chang and T.-M.Yan, Phys.Rev.\textbf{D7} (1973) 1147.

\bibitem{progress}  A.T.Suzuki and A.G.M.Schmidt, Prog.Theor.Phys.\textbf{103
} (2000) 1011.
\end{thebibliography}
\end{document}